# Simultaneous Generation of Arbitrary Assembly of Polarization States with Geometrical-Scaling-Induced Phase Modulation

Ya-Jun Gao,[1,‡] Xiang Xiong,[1,‡] Zhenghan Wang,[1] Fei Chen,[1] Ru-Wen Peng,[1,*] and Mu Wang[1,2,†]

[1]*National Laboratory of Solid State Microstructures, School of Physics, and Collaborative Innovation Center of Advanced Microstructures, Nanjing University, Nanjing 210093, China*
[2]*American Physical Society, Ridge, New York 11961, USA*

Manipulating the polarization of light on the microscale or nanoscale is essential for integrated photonics and quantum optical devices. Nowadays, the metasurface allows one to build on-chip devices that efficiently manipulate the polarization states. However, it remains challenging to generate different types of polarization states simultaneously, which is required to encode information for quantum computing and quantum cryptography applications. By introducing geometrical-scaling-induced (GSI) phase modulations, we demonstrate that an assembly of circularly polarized (CP) and linearly polarized (LP) states can be simultaneously generated by a single metasurface made of $L$-shaped resonators with different geometrical features. Upon illumination, each resonator diffracts the CP state with a certain GSI phase. The interaction of these diffractions leads to the desired output beams, where the polarization state and the propagation direction can be accurately tuned by selecting the geometrical shape, size, and spatial sequence of each resonator in the unit cell. As an example of potential applications, we show that an image can be encoded with different polarization profiles at different diffraction orders and decoded with a polarization analyzer. This approach resolves a challenging problem in integrated optics and is inspiring for on-chip quantum information processing.

## I. INTRODUCTION

As an intrinsic feature of electromagnetic waves, polarization has facilitated numerous applications in photonics and information technology [1–4]. So far, many methods have been employed to modulate the polarization state of light. The most common approach uses optical chirality, where the refractive index differs for the right- and left-handed circular polarized (RCP and LCP) light [5–7]. Birefringence can also transform the polarization state of the incident light based on different phase velocities of two orthogonal components of the electric field [8]. However, these approaches are usually volumetric, so the device has to reach a certain size to tune the polarization state, which is not favorable for integrated photonics. Recently, it has been discovered that the metasurface can effectively manipulate the polarization state of light [9–24]. For the circularly polarized (CP) incidence, a rotation-induced geometrical phase, known as the Pancharatnam-Berry (P-B) phase, is generated by rotating the anisotropic building elements [25–35]. Yet, this phase modulation on LCP and RCP is strongly correlated. More specifically, when the angular rotation of the element is $\theta$, the phase imposed on LCP is $2\theta$ and that on RCP is $-2\theta$ [15]. Macroscopically, the beams with the opposite circular polarization are deflected by an angle with the same value yet opposite sign if the imposed P-B phase on each building element possesses a linear gradient. If the building elements are arbitrarily arranged, the imposed P-B phase does not possess a linear gradient; hence, the output beams become pairs of conjugated CP states, or pairs of conjugated elliptical polarized states, or pairs of LP states with the same polarization [36]. This intrinsic strong correlation of the output states prevents the P-B phase approach from generating an assembly of polarization states of different types [25,28,36].

However, the simultaneous generation of different polarization states is essential for information encoding and quantum cryptography [19,20,37,38]. For example, the paradigm quantum key distribution protocol BB84 employs four of the six polarization states (LCP, RCP, horizontal, vertical, +45°, and −45° LP, respectively) to constitute two bases [39]. The protocol relies on the quantum property that accessing information is only possible at the cost of disturbing the signal if the employed states are nonorthogonal [40]. To meet the challenging requirements of quantum information, a jigsaw puzzle approach has been proposed, which combines six individual regions featuring different P-B phases to generate four LP states and two CP states [29]. However, in addition to the technical issues of the uniformity of output beams and sample size, such a combination approach does not provide a real sense of integration.

*Corresponding author.
rwpeng@nju.edu.cn
†Corresponding author.
muwang@nju.edu.cn
‡Y.-J. G. and X. X. contributed equally to this work.

Here, we report a new strategy to design the metasurface, where each element (resonator) of the metasurface diffracts either a RCP or LCP state, with an additional phase modulation determined by its geometry features. The interaction of these diffracted fields leads to the desired output beams, where the polarization state and the propagation direction can be accurately controlled by the geometrical shape, size, and spatial sequence of each resonator in the unit cell of the metasurface. In contrast to the rotation-induced P-B phase, here the add-on phase on each resonator depends on its geometrical shape and size, so it is a geometrical-scaling-induced (GSI) phase. We demonstrate that multiple beams with different types of polarization states can be simultaneously generated from a single metasurface. The unit cell of the metasurface consists of an assembly of $L$-shaped resonators with different arm length and width, and their symmetrical isomers (mirror images), as illustrated in Fig. 1. Depending on the category and sequence of the resonators in the unit cell, left/right-handed CP states ($|L\rangle/|R\rangle$) and horizontally/vertically/$+45°/-45°$ LP states $[|H\rangle/|V\rangle/(\sqrt{2}/2)(|H\rangle+|V\rangle)/(\sqrt{2}/2)(|H\rangle-|V\rangle)]$ can be simultaneously generated with any desired combination and propagation direction. Furthermore, as an example of potential applications, we demonstrate that an image can be encoded with different polarization profiles at different diffraction orders ($n = \pm 1, \pm 3$) and decoded with a polarization analyzer.

## II. THEORY AND METASURFACE DESIGN

When a plane wave shines on a metasurface made of an array of unit cells ($N_1 \times N_2$), the diffraction field $\vec{E}(k_x, k_y)$ at a far-field point $S$ can be represented by the superposition of each diffraction field induced by the unit cells [41],

$$\vec{E}(k_x, k_y) = e^{i[-\omega t + (2\pi/\lambda)r_0]}$$
$$\times \sum_{u=1}^{N_1}\sum_{v=1}^{N_2} \int \vec{E}(x', y') e^{-i[k_x(x'+x_u)+k_y(y'+y_v)]} dx'dy'$$
$$= e^{i[-\omega t + (2\pi/\lambda)r_0]} \int_{D_x, D_y} \vec{E}(x, y) e^{-i(k_x x + k_y y)} dxdy$$
$$\times \sum_{u=1}^{N_1}\sum_{v=1}^{N_2} e^{-ik_x x_u} e^{-ik_y y_v}, \quad (1)$$

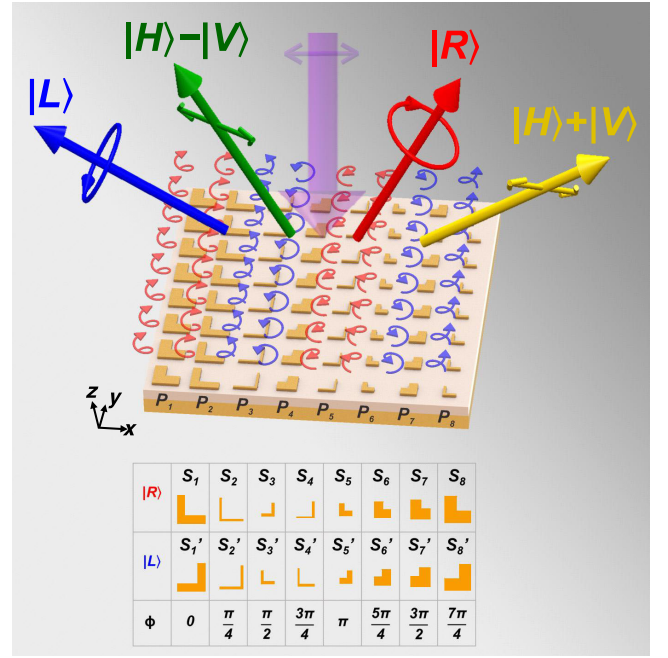

FIG. 1. Schematics showing the generation of multiple coherent beams with a different type of polarization state from a metasurface made of $L$-shaped resonators with different geometrical features. The LP incident beam is represented by the purple arrow. Each resonator diffracts a CP state with a certain GSI phase $\phi$. By controlling the category and the spatial sequence of the $L$-shaped resonators in the unit cell, the desired CP (the red clockwise arrow represents RCP, and the blue counterclockwise one represents LCP) and/or LP (illustrated by the large straight arrows) beams can be simultaneously generated. The resonators $P_j$ ($j = 1, 2, ..., 8$) form the unit cell of the metasurface. The resonators are selected from the pool $\{S_i\}$ and $\{S'_i\}$ ($i = 1, 2, ..., 8$). The resonators in $\{S_i\}$ generate RCP, and those from $\{S'_i\}$ generate LCP, as illustrated in the lower panel.

where $r_0$ is the distance from the center of the metasurface to the point $S$, $\lambda$ is the incident wavelength, $\vec{E}(x, y)$ is the electric field over the unit cell, and $(x_u, y_v)$ are the coordinates of the center of the unit cell. Here, $k_x$ ($k_y$) is the $x$ ($y$) component of the wave vector; $D_x$ ($D_y$) is the periodicity of the unit cell in the $x$ ($y$) direction. By defining $k_x = (2\pi/\lambda)\sin\theta_x = (2\pi n/D_x) = k_n$, $k_y = (2\pi/\lambda)\sin\theta_y = (2\pi m/D_y) = k_m$, where $m$ and $n$ are either positive or negative integers or zero (the diffraction order) and $\theta_x$ and $\theta_y$ are the diffraction angles, the maxima of the diffraction field $\vec{E}(k_x, k_y)$ can be expressed as

$$\vec{E}(k_n, k_m) = e^{i[-\omega t + (2\pi/\lambda)r_0]} N_1 N_2$$
$$\times \int_{D_x, D_y} \vec{E}(x, y) e^{-i(k_n x + k_m y)} dxdy. \quad (2)$$

Equation (2) indicates that the angular spectrum of the metasurface becomes discrete. Now, we consider a unit cell with $Q$ resonators. If the metasurface is designed such that $D_y < \lambda$ is satisfied, there will be no diffraction in the $y$ direction; i.e., $m$ remains zero. Thus, we focus on the contribution of the resonator assembly in the $x$ direction only. Let $\vec{E}_{P_j}$ denote the diffraction field of each resonator ($P_j$ with $j = 1, 2, …, Q$) in the unit cell. It follows that the $n$th order of the diffraction field of the metasurface is expressed as

$$\vec{E}(k_n) = e^{i[-\omega t+(2\pi/\lambda)r_0]} N_1 \sum_{j=1}^{Q} \int_{(j-\frac{Q}{2}-1)\frac{D_x}{Q}}^{(j-\frac{Q}{2})\frac{D_x}{Q}} \vec{E}_{P_j} e^{-i(2\pi n/D_x)x} dx. \tag{3}$$

To generate multiple beams with desired CP and/or LP states, the amplitude and the phase of the diffracted light from each resonator in the unit cell should be elaborately designed.

In this article, the resonators in the unit cell are the $L$-shaped gold structures sitting on top of a $SiO_2$-gold-silicon substrate. The incident light shines on the metasurface in the normal direction. The gold layer in the substrate ensures that there is no transmission, and the normalized reflective diffraction intensity of the resonator is enhanced. Physically, because of the existence of the gold layer, the diffraction from the resonator is the interference of the irradiation of the $L$-shaped structure and the reflection field of the gold layer. Meanwhile, the reflection field is composed of the mirror image of the incident light and the irradiation of the mirror image of the $L$-shaped resonators [11]. It follows that $\vec{E}_{P_j}$ in Eq. (3) can be expressed as $-\vec{E}_{inc} + \vec{E}_{rad,j}[-e^{i(2\pi h_1/\lambda)} + e^{-i(2\pi h_1/\lambda)}]$. Here, for simplicity of the expression, we take the refractive index of the $SiO_2$ layer to be 1. The details of the treatment of a real $SiO_2$ layer are provided in the Supplemental Material of Ref. [11]. Here, $\vec{E}_{inc}$ is the electric field of the incident light, $\vec{E}_{rad,j}$ is the irradiation field of the $L$-shaped structure, $\lambda$ is the wavelength, and $h_1$ is the thickness of the $SiO_2$ layer. Note that $\vec{E}_{rad,j}$ stands for the irradiation capability of the $L$-shaped structure, and it is related to the geometrical features of the $L$ pattern. It follows that $\vec{E}_{P_j}$ depends on the incident wavelength, the thickness of the $SiO_2$ layer in the substrate, and the geometrical size of the $L$ pattern. By carefully selecting these parameters, the phase difference between the $x$ and $y$ components of $\vec{E}_{P_j}$ can be adjusted to 90° or 270°. In this way, a CP state is generated. Most importantly, by choosing a certain length and width of the arms of the $L$ resonator, a specific GSI phase ($\phi$) is imposed on the generated CP state. Moreover, by taking the mirror image of the $L$ resonator (with the opening pointing in the 135° direction), the CP state of the opposite handedness is generated [36]. Practically, we construct a pool of 16 resonators, where eight resonators make a set $\{S_i\}$ ($S_1, S_2,...,S_8$); the other eight are their mirror-image structures $\{S_i'\}$ ($S_1', S_2',...,S_8'$), as illustrated in the lower panel of Fig. 1. The geometrical sizes of the $L$-shaped resonators are listed in the Supplemental Material [36]. The resonators in the unit cell are selected from the pool of $\{S_i\}$ and $\{S_i'\}$.

Figure 2 shows the amplitude and the phase characteristics of the diffraction from each resonator in $\{S_i\}$ and $\{S_i'\}$ simulated by the finite difference time domain (FDTD) method. The $x$-polarized incident light propagates in the $-z$ direction, and the wavelength is set as 1300 nm. The center-to-center separation of the neighboring resonators in the $x$ direction is 700 nm. We focus on three parameters of each resonator: the reflectance ratio of the $y$ component and the $x$ component ($r_{yx}/r_{xx}$), the reflected phase difference between the $y$ and $x$ components ($\Delta\phi$), and the GSI phase $\phi$ of the resonator. Here, $r_{yx}/r_{xx}$ and $\Delta\phi$ characterize the polarization state of the diffraction from the resonator. For the CP state, the reflectance ratio equals 1, and the phase difference equals 90° or 270°. Figures 2(a) and 2(b) illustrate the simulated reflectance ratio $r_{yx}/r_{xx}$, and Figs. 2(c) and 2(d) show the phase difference between the $y$ and $x$ components when the incident light shines on the resonators in $\{S_i\}$ and $\{S_i'\}$, respectively. When the resonator is selected from $\{S_i\}$, the reflectance ratio remains unity, and the phase of the $x$ component is 90° advanced, indicating that the resonators in $\{S_i\}$ generate a RCP state. For the resonators in $\{S'_i\}$, the reflectance ratio remains unity, and the phase of the $y$ component is 90° advanced, indicating that the resonators in $\{S'_i\}$ generate a LCP state. Figures 2(e) and 2(f) show the simulated GSI phase $\phi$ of the scattered CP state from the 16 resonators. It can be seen that the range of $\phi$ for the resonators in $\{S_i\}$ covers 0 to $2\pi$ with a step of $\pi/4$, and the same applies for that in $\{S'_i\}$. The simulation results in Fig. 2 indicate that upon normal illumination of an $x$-polarized incidence, resonators in $\{S_i\}$ ($\{S'_i\}$) generate a RCP (LCP) state, and a GSI phase $\phi$ is added to the diffraction field of each resonator.

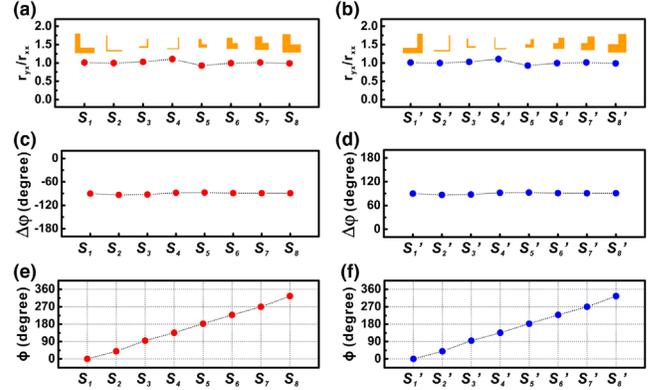

FIG. 2. Simulation results of 16 $L$-shaped resonators ($S_1, S_2, …, S_8$ and $S_1', S_2', …, S_8'$), where the $x$-polarized incident light propagates in the $-z$ direction, and the wavelength is 1300 nm. (a) The simulated reflectance ratio of the $y$ component and the $x$ component of eight resonators in $\{S_i\}$ ($i = 1, 2, …, 8$). (b) The simulated reflectance ratio of the $y$ component to the $x$ component of eight resonators in $\{S_i'\}$ ($i = 1, 2, …, 8$). (c) The simulated reflected phase difference between the $y$ and $x$ components of eight resonators in $\{S_i\}$ ($i = 1, 2, …, 8$). (d) The simulated reflected phase difference between the $y$ and $x$ components of eight resonators in $\{S_i'\}$ ($i = 1, 2, …, 8$). (e) The simulated GSI phase $\phi$ added to the diffracted state from the elements in $\{S_i\}$ ($i = 1, 2, …, 8$). (f) The simulated GSI phase $\phi$ added to the diffracted state from the elements in $\{S_i'\}$ ($i = 1, 2, …, 8$).

The number of diffraction beams from the metasurface is determined by the diffraction order $n$, which depends on the ratio of the periodicity of the unit cell in the $x$ direction, $D_x$, and the wavelength $\lambda$ [$|n| \leq (D_x/\lambda)$]. Currently, the separation of resonators in the $x$ direction is 700 nm, and the incident wavelength is 1300 nm. To generate at least four diffraction beams, $n$ should satisfy $|n| \geq 2$. It follows that the number of resonators within the unit cell should be at least 4. On the other hand, it is known that the higher angle diffraction usually gets lower diffraction efficiency, which is defined as the ratio of the intensity of the diffracted beam to that of the incident beam. For easier selection of the diffraction beams with higher efficiency, we introduce eight resonators in the unit cell ($Q = 8$). Meanwhile, the assembly of the resonators within the unit cell may have $16^8$ combinations in total. The combination can be reduced by considering only even (or odd) diffraction orders. More specifically, if the diffraction fields $\vec{E}_{P_j}$ and $\vec{E}_{P_{j+4}}$ ($j = 1, 2, 3, 4$) possess a phase difference $\pi$, according to Eq. (3), the output beams only have odd orders, which means that only four resonators ($P_1$, $P_2$, $P_3$, and $P_4$) are essentially independent. Similarly, if $\vec{E}_{P_j}$ and $\vec{E}_{P_{j+4}}$ ($j = 1, 2, 3, 4$) are in the same phase ($2\pi$), the output beams will be in even orders. In both cases, the combinations of the resonators are reduced sharply from $16^8$ to $16^4$.

To elucidate how the desired state is formed, we first select eight resonators, all from $\{S_i\}$ ($i = 1, 2, …, 8$). The resonators are arranged such that a linear GSI phase gradient is established, as shown in Fig. 3(a). According to Eq. (3), the polarization state of the first diffraction order is

$$\vec{E}(k_1) = N_1 \sum_{j=1}^{8} \int_{(j-5)\frac{D_x}{8}}^{(j-4)\frac{D_x}{8}} \vec{E}_{P_j} e^{-i(2\pi/D_x)x} dx$$
$$= a_1 \sum_{j=1}^{8} \vec{E}_{P_j} e^{-i(\pi/4)(j-4.5)} = 8a_1 e^{i(7/8)\pi}|R\rangle, \quad (4)$$

where $a_1$ is a parameter related to the wavelength and periodicity in the $x$ direction. It is noteworthy that for the scenario of Fig. 3(a), $n = 1$ is the only diffraction beam coming out of the metasurface, which is a pure RCP state. If $n$ takes a value other than 1, according to Eq. (3), $\vec{E}(k_n) = 0$ ($n \neq 1$). When the resonators in the unit cell are all taken from $\{S'_i\}$ in a similar way, i.e., $S'_1, S'_2, S'_3, …, S'_8$, a pure LCP state is generated at the first order ($n = 1$).

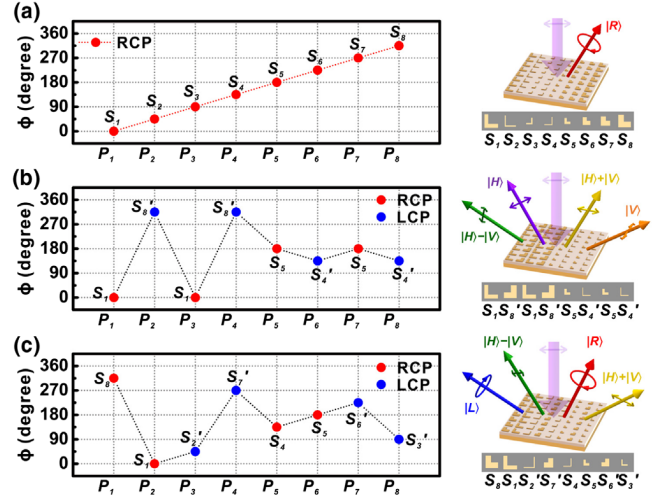

FIG. 3. GSI phase $\phi$ and the handedness (represented by the red or blue dots) of each resonator in the unit cell. The metasurface and the unit cell are illustrated on the right. In the unit cell, the geometrical features of each resonator vary. (a) $\phi$ of eight resonators all selected from $\{S_i\}$ with a linear phase gradient. The output is an RCP state. (b) $\phi$ of eight resonators, four from $\{S_i\}$ and four from $\{S'_i\}$. The metasurface generates four LP states. (c) $\phi$ of another set of eight resonators selected from $\{S_i\}$ and $\{S'_i\}$, respectively. The output is two CP and two LP states.

It is known that a LP state can be superimposed by LCP and RCP with the same amplitude. To diffract a LP beam from the metasurface with the resonators generating CP states, $P_1, P_2, P_3$, and $P_4$ should be selected from both $\{S_i\}$ and $\{S'_i\}$. If $P_1, P_2, P_3$, and $P_4$ are selected from the same set only (i.e., either from $\{S_i\}$ or $\{S'_i\}$), this combination will be 8192 ($C_2^1 8^4$). Therefore, when the resonators are selected from both $\{S_i\}$ and $\{S'_i\}$, the combination of the resonators is reduced from 65 536 ($16^4$) to 57 344 ($16^4 - C_2^1 8^4$). Furthermore, since the unit cell is periodically arranged in the metasurface, once the resonators within the unit cell are sequentially permutated, macroscopically the output states resume. Thus, the number of independent combinations will be 1/8 of 57 344. Let us take the sequence $S_1, S_2, S_3, S_4, S_5, S_6, S_7, S_8$ as an example. Considering sequential permutation of the elements in the unit cell, seven other sequences generate the same output states, such as $S_2, S_3, S_4, S_5, S_6, S_7, S_8, S_1$ and $S_3, S_4, S_5, S_6, S_7, S_8, S_1, S_2$ and $S_4, S_5, S_6, S_7, S_8, S_1, S_2, S_3$, etc. It follows that the number of independent combinations is reduced to 7168 (57 344/8). Thereafter, to generate multiple beams with the desired CP and/or LP states, the resonators within the unit cell should be elaborately selected via a parameter scan of the 7168 combinations based on Eq. (3).

Figure 3(b) illustrates the scenario when the resonators are selected as $S_1$, $S'_8$, $S_1$, $S'_8$, $S_5$, $S'_4$, $S_5$, $S'_4$. The corresponding phases and handedness of the diffracted states from each resonator are $|R\rangle$, $e^{i(7\pi/4)}|L\rangle$, $|R\rangle$, $e^{i(7\pi/4)}|L\rangle$, $e^{i\pi}|R\rangle$, $e^{i(3\pi/4)}|L\rangle$, $e^{i\pi}|R\rangle$, and $e^{i(3\pi/4)}|L\rangle$, respectively. Here, $n$ satisfies $|n| \leq (D_x/\lambda)$ due to the restriction $|\sin\theta_x| \leq 1$. Since the separation of the resonators in the $x$ direction is 700 nm, $D_x$ becomes 5600 nm. The incident wavelength is 1300 nm. It follows that $|n| \leq 4$. Furthermore, since $\vec{E}_{P_j}$ and $\vec{E}_{P_{j+4}}$ ($j = 1, 2, 3, 4$) possess a phase difference of $\pi$, the observable $n$ can only be $\pm 1, \pm 3$. In Eq. (3), by setting $n = 1$, the polarization state of the first-order diffraction is $(\sqrt{2}/2)e^{i(3\pi/8)}(|H\rangle + |V\rangle)$. By taking $n = -1, +3, -3$ in Eq. (3), the corresponding polarization states become $e^{-i(5\pi/8)}|H\rangle$, $e^{i(3\pi/8)}|V\rangle$, and $(\sqrt{2}/2)e^{-i(5\pi/8)}(|H\rangle - |V\rangle)$, respectively. In this way, four coherent LP states are generated.

By selecting the geometrical shape, size, and spatial sequence of the resonators in the unit cell, we can obtain any type of polarization state propagating in the desired direction. As illustrated in Fig. 3(c), if the unit cell is selected as $S_8$, $S_1$, $S'_2$, $S'_7$, $S_4$, $S_5$, $S'_6$, $S'_3$, the corresponding phase and handedness of the diffraction from these eight resonators are $e^{i(7\pi/4)}|R\rangle$, $|R\rangle$, $e^{i(\pi/4)}|L\rangle$, $e^{i(6\pi/4)}|L\rangle$, $e^{i(3\pi/4)}|R\rangle$, $e^{i\pi}|R\rangle$, $e^{i(5\pi/4)}|L\rangle$, $e^{i(2\pi/4)}|L\rangle$, respectively. For the aforementioned reason, $n$ can only be $\pm 1$ and $\pm 3$. It follows that, corresponding to $n = 3, 1, -1, -3$, the polarization states become $(\sqrt{2}/2)e^{-i(5/8)\pi}(|H\rangle + |V\rangle)$, $e^{i(5/8)\pi}|R\rangle$, $(\sqrt{2}/2)e^{-i(1/8)\pi}(|H\rangle - |V\rangle)$, and $e^{i(9/8)\pi}|L\rangle$, respectively. This result means that two LP ($+45°$, $-45°$) states and two CP (RCP, LCP) states with fixed phase differences are simultaneously generated.

## III. EXPERIMENTAL

To verify the theoretical design, we fabricate two arrays of $L$-shaped resonator assemblies on top of a $SiO_2$-gold-silicon substrate: One scenario generates four LP states ($+45°$-LP, $-45°$-LP, horizontally LP, and vertically LP), and the other generates two LP ($+45°$-LP, $-45°$-LP) states and two CP (RCP, LCP) states, as shown in Figs. 4(a) and 5(a), respectively. To characterize these samples, a supercontinuum laser and a Glan-Taylor polarizer are used to generate the LP incident beam. The propagation direction of each diffraction beam satisfies $\sin\theta_x = (n/D_x)\lambda$. An angle-resolved detector measures the diffraction intensity in the ranges of $-60°$ to $-12°$ and $12°$ to $60°$. The region between $12°$ and $-12°$ which is marked as the area between two white dash lines in Figs. 4(b) and 5(b), cannot be detected due to the technical restriction of the reflection mode. To identify the polarization state of each diffracted beam, an achromatic quarter-wave plate and/or a polarizer are installed in front of the detector (see Ref. [36] for details). The angle-resolved diffraction spectra are illustrated in Figs. 4(b) and 5(b). Experimentally, we define a parameter, the light intensity ratio, to characterize the purity of each generated state. For the LP beam, the light intensity ratio is defined as the ratio of the intensity of the investigated beam and the intensity detected with the orthogonal polarization; for the CP beam, it is defined as the ratio of the intensity of the investigated beam and that with the conjugated CP. For the ideal situation, this ratio should be infinity for a pure CP/LP state, which requires that each resonator in the unit cell possesses the uniform diffraction intensity and the elaborately designed phase. In reality, however, the intensity of the diffraction field from different resonators varies. In selecting the resonators for $\{S_i\}$ and $\{S'_i\}$, we practically choose the resonators with the diffraction intensity within the range of $0.85 \pm 0.05$ to guarantee that the intensity ratio is large.

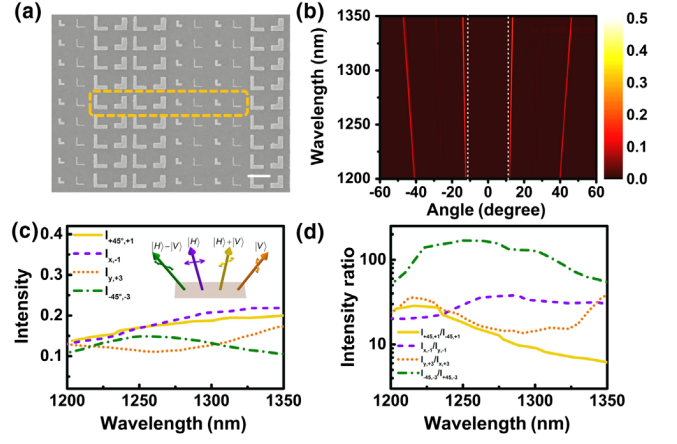

FIG. 4. (a) SEM micrograph of the fabricated metasurface as designed in Fig. 3(b). The unit cell is marked by the dashed-line box. The bar represents 1 $\mu$m. (b) Measured angle-resolved diffraction spectra. The four bright lines represent the diffracted beams measured at a different wavelength at different diffraction angles. The color bar stands for the intensity. (c) Normalized beam intensity of four diffracted beams. Note that $I_{\text{pol,num}}$ represents the intensity of "numth" order, with "pol" polarization. In the wavelength range of 1200–1350 nm, the normalized intensity of each diffraction beam is higher than 10%. At 1300 nm, the intensity of each diffraction order is higher than 12%, and the total normalized diffraction intensity reaches 65%. (d) Measured light intensity ratio of each beam, which is defined as the intensity of the measured LP (CP) beam divided by the intensity of its orthogonal (conjugated) beam. The intensity ratio of each beam is larger than 10 for the wavelength between 1200 nm and 1275 nm, and the maximum reaches 168 at 1260 nm.

For the sample shown in Fig. 4(a), four diffraction beams, from left to right, correspond to negative third, negative first, first, and third order of diffraction, have been detected [Fig. 4(b)]. To characterize the diffraction efficiency, we define the normalized beam intensity as the ratio of the intensity of the diffracted beam and that of the incident beam, as illustrated in Fig. 4(c). In the wavelength range of 1200–1350 nm, the normalized intensity of each diffraction beam is higher than 10%, and the total normalized intensity of diffraction beams is higher than 50%. In particular, at 1300 nm, the intensity of each diffraction order is higher than 12%, and the total normalized diffraction intensity reaches 65%. Figure 4(d) shows the light intensity ratio of the four beams. It can be seen that the intensity ratio of each beam is larger than 10 for the wavelength between 1200 nm and 1275 nm, and the maximum reaches 168 at 1260 nm. Figure 4(d) therefore confirms that the polarization states of the output beams are in agreement with theoretical expectations.

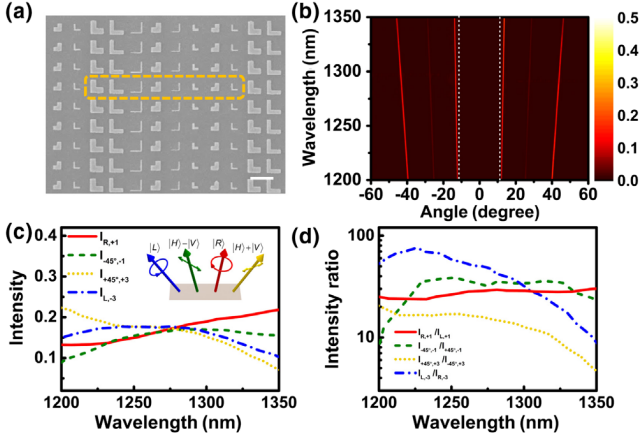

FIG. 5. (a) SEM micrograph of the fabricated metasurface as designed in Fig. 3(c). The unit cell is marked by the dashed-line box. The bar stands for 1 $\mu$m. (b) Measured angle-resolved diffraction spectra. Four bright lines represent the diffracted beams measured at a different wavelength at different diffraction angles. The color bar stands for the intensity. (c) Normalized beam intensity of four diffracted beams. Here, $I_{\text{pol,num}}$ represents the intensity of "numth" order with "pol" polarization. Within the wavelength range of 1200–1325 nm, the normalized intensity of each diffracted beam is higher than 8%. At 1275 nm, the intensity of each diffraction beam is higher than 16%, and the total normalized intensity reaches 68%. (d) Measured light intensity ratio of each diffracted beam, which is defined as the intensity of the measured LP (CP) beam divided by the intensity of its orthogonal (conjugated) beam. The intensity ratio of each beam is larger than 10 for the wavelength between 1210 nm and 1300 nm, and the maximum reaches 76 at 1225 nm.

To generate two CP and two LP beams, we fabricate the structures shown in Fig. 5(a). One can identify that the resonators in the unit cell are $S_8$, $S_1$, $S'_2$, $S'_7$, $S_4$, $S_5$, $S'_6$, $S'_3$, respectively. Figure 5(b) indicates that four diffraction beams are simultaneously generated, and the normalized beam intensity is illustrated in Fig. 5(c). Within the wavelength range of 1200–1325 nm, the normalized intensity of each diffracted beam is higher than 8%, and the total normalized intensity in these four directions is larger than 58%. At 1275 nm, the intensity of each diffraction beam is higher than 16%, and the total normalized intensity reaches 68%. The light intensity ratio of these four beams is presented in Fig. 5(d), which is always higher than 10 in the wavelength range 1210–1300 nm, and the maximum reaches 76 at 1225 nm. Figure 5(d) confirms that two CP states and two LP states have indeed been experimentally realized.

The polarization state can be applied to encode information in cryptography [18,32]. Here, we experimentally demonstrate encoding of information with different polarization profiles at different diffraction orders ($n = \pm 1, \pm 3$) and decoding the information with an analyzer. A pattern of an overlapped circle and square is selected for encoding. The resonators are selected as $S_8$, $S_1$, $S'_2$, $S'_7$, $S_4$, $S_5$, $S'_6$, and $S'_3$. Three unit cells with different sequences—{$S1$} ($S_8$, $S_1$, $S'_2$, $S'_7$, $S_4$, $S_5$, $S'_6$, $S'_3$), {$S2$} ($S_1$, $S_8$, $S'_3$, $S'_6$, $S_5$, $S_4$, $S'_7$, $S'_2$), and {$S3$} ($S_1$, $S_4$, $S'_3$, $S'_2$, $S_5$, $S_8$, $S'_7$, $S'_6$)—are applied to construct the metasurface in regions 1, 2, and 3, respectively, as shown in Figs. 6(a) and 6(b). In each region, the unit cell is reproduced with a periodicity of 5600 nm in the $x$ direction and 700 nm in the $y$ direction. By shining an $x$-polarized incident light on the sample, different polarization-state profiles appear at the different diffraction orders, as shown in Fig. 6(b), which can be regarded as an encoding process. For example, in Fig. 6(b), at the negative third order, the polarization states generated in the regions 1, 2, and 3 are LCP, 45°-LP, and −45°-LP, respectively. For the negative first order, however, the polarization states become −45°-LP, RCP, and LCP instead. If an analyzer is placed in front of the charge-coupled device (CCD) camera, the encoded pattern of the overlapped circle and square can be decoded by displaying different contrast distributions at different diffraction orders.

We experimentally fabricate the sample shown in Figs. 6(c) and 6(d), where the whole structure is 649.6 × 448 $\mu$m in size. A LED light source combined with a Glan-Taylor polarizer generates the $x$-polarized incident light. A 10-times objective is mounted in front of the CCD camera to visualize the diffraction image at each order. To decode the information, either a linear polarizer or a circular polarizer is placed in front of the CCD camera. Upon illumination of the incident light, a combination of four polarization states (−45°-LP, 45°-LP, RCP, and LCP) is generated. Accordingly, we can select one of four polarization states (−45°-LP, 45°-LP, RCP, and LCP) as the analyzers, as shown in Fig. 6(e). For different analyzers, the contrast pattern at different diffraction orders varies. The bright contrast indicates that the polarization of the diffracted beam in the region is identical to that of the analyzer. The dark contrast, on the other hand, indicates that the polarization of the diffracted beam in the local area is either orthogonal to that of the linear analyzer or in the conjugated state as that of the circular analyzer. The grey contrast means that the polarization of the diffracted beam is not the same type of polarization state as the analyzer;

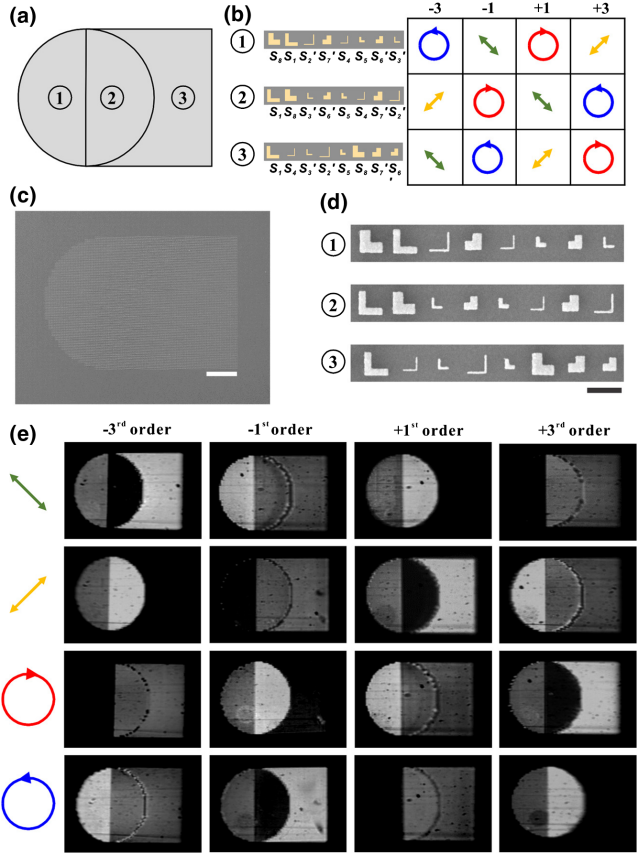

FIG. 6. (a) Schematic of the overlapped circle and square, where three regions (1, 2, and 3) can be identified. These regions are selected for polarization encoding. (b) Morphology of three types of unit cells {S1}, {S2}, and {S3} and the corresponding polarization states appearing at diffraction orders −3, −1, +1, and +3, respectively. Three different sequences are designated to regions 1, 2, and 3, respectively. The incident beam is $x$ polarized. The green and yellow straight arrows represent the −45°-LP and +45°-LP states. The blue counterclockwise (red clockwise) arrow represents the LCP (RCP) state. (c) Micrograph of the fabricated pattern. The bar represents 100 $\mu$m. (d) SEM micrograph of the unit cell designated to regions 1, 2, and 3, respectively. The bar represents 1 $\mu$m. (e) Diffraction images at orders −3, −1, +1, and +3 visualized by the CCD camera. Four types of analyzers are put in front of the CCD, respectively, which is represented by the colored symbols on the left side. Two linear polarized analyzers are oriented at the −45° (green arrow) and 45° (yellow arrow) directions, and two circular polarized analyzers are for RCP (red circle) and LCP (blue circle), respectively. Three different shades —bright, dark, and grey—appear at different regions depending on the selection of the analyzer.

i.e., for the LP analyzer, the polarization of the diffraction is CP; for a CP analyzer, the diffracted polarization is LP. In this way, the polarization state can be identified from the contrast distribution at each diffraction order with the help of an analyzer. Figure 6 demonstrates that the information can be encoded with multiple polarization states and decoded by an analyzer.

## IV. DISCUSSION

When the metasurface is introduced, the interface is no longer homogeneous. Meanwhile, conventional Snell's law has to be modified: Depending on the phase distribution on the interface, the diffracted beams can essentially be directed to any orientation [9,28,29]. The propagation direction and the number of diffraction beams can be accurately modulated. The propagation direction of the diffraction beam satisfies $\sin\theta_x = (n\lambda/D_x)$. The number of beams is decided by both the diffraction order $n$ and the combination of the resonators within the unit cell. More specifically, the diffraction order $n$ follows $|n| \leq (D_x/\lambda)$. Meanwhile, the combination of the resonators within the unit cell can lead to the extinction of the electric field in certain diffraction orders. For example, when the resonators in the unit cell are arranged as $S_1, S_2, \ldots, S_8$ [Fig. 3(a)], the electric field vanishes in the other diffraction orders, except for the first order. In the Supplemental Material [36], we show another combination of resonators where both orders $n = \pm 1$ are generated. As illustrated in Figs. 3(b) and 3(c), $n = \pm 1, \pm 3$ are diffracted from the metasurface to four different directions. Therefore, by tuning $D_x$, $\lambda$, and the combinations of the resonators within the unit cell, the number of diffraction beams and the propagation directions can be tuned.

Once the resonators in the unit cell have been selected, the sequence of these resonators determines which polarization state comes from which diffraction order. Let us take the unit cell consisting of $S_8, S_1, S'_2, S'_7, S_4, S_5, S'_6$, and $S'_3$ as an example, which contains the same group of resonators as that shown in Fig. 5(a). Consider the special scenario in which the eight resonators within the unit cell can only be selected from $S_8, S_1, S'_2, S'_7, S_4, S_5, S'_6$, and $S'_3$. There are 384 ($C_8^1 C_6^1 C_4^1 C_2^1$) sequences satisfying that resonators ($P_1$, $P_2$, $P_3$, and $P_4$) are independent, and meanwhile, the diffraction fields $\vec{E}_{P_j}$ and $\vec{E}_{P_{j+4}}$ ($j = 1, 2, 3, 4$) possess a phase difference $\pi$. Thus, the output beams of the 384 sequences only have odd orders. Moreover, considering the sequential-permutation-induced degeneracy, the number of real independent sequences will be 1/8 of 384 (48). Eight of these 48 combinations can generate two CP (LCP, RCP) states and two LP (+45°, −45°) states. For example, if the sequence is rearranged as $S_4, S_1, S'_2, S'_3, S_8, S_5, S'_6, S'_7$, the output states will be RCP, −45°-LP, LCP, and +45°-LP, corresponding to the diffraction orders −3, −1, +1, and +3, respectively. However, if the sequence becomes $S_8, S_1, S'_6, S'_3, S_4, S_5, S'_2, S'_7$, the output states turn out to be LCP, +45°-LP, RCP, and −45°-LP instead. More examples are provided in Table S3 in the Supplemental Material [36].

The generation of four (or more) different polarization states is required by the paradigm quantum key distribution protocol BB84 [39]. We suggest that the encoding and decoding process in Fig. 6 is similar to the secret key distribution in quantum cryptography via the polarization of photons [39]. In the information cryptography, the information sender (Alice) and the receiver (Bob) share a key, which is used by Alice (Bob) to encrypt (decrypt) the information. Immediately sensing the existence of eavesdroppers in key distribution, which is guaranteed by the principle that quantum measurements cannot occur without perturbing the system [40], ensures the security of key distribution and hence the confidentiality of the information. We suggest that the metasurface in Fig. 5 and the encoding or decoding processes in Fig. 6 can be applied for secure key distribution in quantum information. More specifically, suppose that the key is composed of a sequence of 0 or 1. The metasurface shown in Fig. 5 is placed after a single-photon source. The diffracted photon is in one of the LCP, RCP, $+45°$-LP, and $-45°$-LP states. Here, we designate the binary value 0 as LCP or $+45°$-LP states and the value 1 as the states RCP or $-45°$-LP. It is noteworthy that two nonorthogonal states represent the same value (either 0 or 1), and this scenario strengthens the security of the key [39]. In the case that the system is eavesdropped, the measurement made by the eavesdropper will randomly modify the polarization of the photon. It follows that the error rate of the key, which is defined as the ratio of the number of wrong values to the number of the whole transmitted photons, suddenly increases. The increase of the error rate suggests the existence of eavesdropper, hence the total key will be dropped. After Alice transfers a single photon with the specific polarization one after the other, Bob measures the photon with a randomly chosen analyzer. There are four possible choices for the analyzer, as illustrated in Fig. 6(e). Details about the process of the key transmission are provided in the Supplemental Material [36]. Next, Alice encrypts the information with the shared secret key by a bitwise operation, such as XOR, a logical operation that the value of the output takes 1 only when the values of the binary inputs differ. Bob decrypts the information with the same key with an inverse operation [42]. In this way, the encrypted information is transmitted safely from Alice to Bob. In this process, the key distribution is accomplished with the metasurface and four optical analyzers. This approach provides new perspectives in generating an assembly of nonorthogonal polarization states that are lightweight and have a high integration degree; hence, it is enlightening in developing portable quantum cryptography.

## V. CONCLUSIONS

Encoding, transmission, and processing of information are important issues in quantum information science [42–44]. As the basic unit of quantum information processing, the qubit gate can be constructed with polarization beam splitters, phase shifters, and wave plates, etc., which are bulky and heavy if conventional optical devices are used [43]. Developing new principle optical devices is essential for future integrated photonics and information technology. The approach demonstrated in this article provides a promising solution to generate an arbitrary combination of CP and LP states simultaneously. By judiciously selecting the geometrical size and the symmetry of each resonator in the unit cell and by carefully designing the lattice parameters, we are able to accurately control the output polarization states, the number of output beams, and the propagation direction of each beam. In contrast to the P-B phase modulation, which generates either CP states or LP states, respectively, our approach allows us to form an arbitrary assembly of different polarization states with high integration based on the concept of the GSI phase. We anticipate that such a capability is very supportive in constructing quantum states [20], realizing quantum entanglement [19] and qubit gates [44], and developing quantum cryptography [37,39].

## ACKNOWLEDGMENTS

This work was supported by projects from the National Key R&D Program of China (Grant No. 2017YFA0303702), and the National Natural Science Foundation of China (Grants No. 11634005, No. 11674155, No. 11974177, and No. 61975078).